\documentclass[11pt,a4paper,english]{article}
\pdfoutput=1
\pdfoutput=1
\pdfoutput=1
\pdfoutput=1
\pdfoutput=1

\usepackage{graphicx}
\usepackage{amsmath,amssymb}
\usepackage{babel}

\usepackage[unicode,hidelinks,pdfauthor={A. Penttila},bookmarksopen=true,pdfpagemode=UseNone,pdfstartview={XYZ null null 1.0}]{hyperref}

\usepackage[numbers,sort&compress]{natbib}

\usepackage{lastpage}
\usepackage{fancyhdr}
\fancypagestyle{plain}{

\fancyhead[L]{}
\fancyhead[R]{}
\fancyhead[C]{\small A. Penttil\"a. Quasi-specular reflection from particulate media. JQSRT, in press.}
\fancyfoot{}
\fancyfoot[R]{\small page \thepage/\pageref{LastPage}}
}
\fancypagestyle{empty}{

\fancyhead{}
\fancyfoot{}
\fancyfoot[R]{\small page \thepage/\pageref{LastPage}}
}

\usepackage{textcomp}
\usepackage[utf8]{inputenc}
\usepackage[T1]{fontenc}
\usepackage{tgadventor} 
\usepackage{tgpagella} 
\usepackage[bf,sf]{titlesec}
\def \micron {{\textmu}m}

\newcommand{\degr}[0]{\ensuremath{{}^{\circ}}}

\usepackage{authblk}

\title{Quasi-specular reflection from particulate media}

\author{Antti Penttil\"a}

\affil{Department of Physics, University of Helsinki, P.O. Box 64, 00014 Helsinki University, Finland (Antti.I.Penttila@helsinki.fi)}

\begin{document}

\maketitle
\thispagestyle{empty}

\hspace{10mm}\\
\begin{abstract}
\noindent Specular reflection is known to play an important role in many fields of scattering applications, e.g., in remote sensing, computer graphics, optimization of visual appearance of industrial products. Usually it can be assumed that the object has a solid surface and that the properties of the surface will dictate the behavior of the specular component. In this study I will show that media consisting of wavelength-sized particles can also have a quasi-specular reflection in cases where there is ordered structure in the media. I will also show that the quasi-specular reflection in particulate media is more than just a surface effect, and planar particle arrangement below the very surface can give arise to quasi-specular reflection. This study shows that the quasi-specular reflection may contribute in some cases in the backscattering direction, together with coherent backscattering and shadow-hiding effects.
\end{abstract}

\hspace{20mm}\\
Cite as: Antti Penttil\"a. Quasi-specular reflection from particulate media. JQSRT, in press, DOI: 10.1016/j.jqsrt.2013.03.007.\\

\clearpage
\pagestyle{plain}

\section*{Introduction}

Specular reflection is one of the first topics taught in optics. In an idealized case of perfectly flat solid interface between two materials the geometrical optics approach is that one part of the incident light (or other wavelengths) is reflected into the specular direction and the rest is refracted into the second material. The amount of reflected and refracted power is given by the Fresnel coefficients. The approach that is often used to describe scattering from surfaces that are not ideally flat is that there is a diffuse component and a specular component in the scattered signal. This approximation is based on the microfacet idea \cite{Torrance-1967-specular,Beckmann-1963-surfaces} where it is assumed that the surface constitutes of small planar facets. The facets are thought to be small, but still significantly larger than the wavelength so that their scattering can be modeled by specular reflection to the direction that is governed by the local normal vector of the facet. The specular reflection from surface consisting of microfacets is the sum of the first-order specular reflections from suitable facets, i.e., from facets where the local normal is such that the incidence angle, observation angle and the local normal will form a specular geometry. The more rough the surface is, the more the local normals are deviated from the average normal, and the more the specular signal is decreased and spread around the average specular direction. The diffuse scattering from microfacets comes from the multiple-scattered signals, but this alone cannot explain well the diffuse part so we will need some subsurface scattering assumption to explain the total diffuse part.

The specular reflection is modeled and measured in various fields. In remote sensing it is often called a 'glint' \cite{Borovoi-2012-Glints} and is observed when measuring scattering from oceans, ice covered areas and cirrus clouds, for example. In astronomy it seems that the radar reflectance properties of some Solar system targets, at least the Saturn moon Titan, need to be explained with a specular component \cite{Janssen-2011-Titan,Soderblom-2012-specular}. In computer graphics reflectance models that will produce specularity are needed to create realistic images. With industrial materials and applications the specular part is usually called 'gloss', and can be either preferred or unpreferred feature in the visual appearance of the product. Certain cosmetic products, paints and papers seek to have high gloss.

In many cases the 'surface approach', connecting the specular reflection to the properties of the geometry of a solid surface, is reasonable and produces good results. The abovementioned oceans and ice covered surfaces, for example, are cases where it is very natural to model the target as a surface that can be divided into small planar facets. With other, clearly particulate targets such as cirrus clouds, it can be argued that the particles (cirrus ice crystals) have planar facets that can act as mirrors. Since the ice crystals in cirrus can have sizes down to micrometer scale it might be a bit precarious to describe their facets as small mirrors, as they start to approach the wavelength size range (in visual range). Nevertheless, it seems that for example the Bennet-Porteus approximation for the strength of the specular reflection from surfaces with wavelength-scale roughness is quite accurate \cite{Penttila-2010-gloss}.

\section*{Quasi-specular reflection}

Industrial materials where specularity is needed often use pigments for surface finish. Pigments are quite small, usually mineral particles. The sizes can vary, but for example with high-gloss paper products the coating pigments are typically in the size range of 1--10 \micron. With this small particles it is quite difficult to claim that parts of the particles surface can act as a microfacets, since these parts can easily be smaller than the wavelength. Actually, the high specularity in paper products is achieved through calendering the paper with high pressure, thus organizing and packing the coating pigments into a very smooth layer. Even though the layer is smooth in macroscopic scale, in micrometer scale it is still constituted of particles and there is void space between the particles for ink to penetrate to the structure when printing.

It seems that the 'surface approach' is not reasonable in cases where we clearly have an interface constituting of wavelength-scale particles. Very glossy papers, for example, can be produced using particulate media so particulate media can produce high reflectance in the specular direction. To distinct this 'glint' or 'gloss' or 'specular-type peak' from the traditional specular reflectance from flat solid surface I will call this effect as quasi-specular reflection (QSR).

\begin{figure}
\centering
\includegraphics{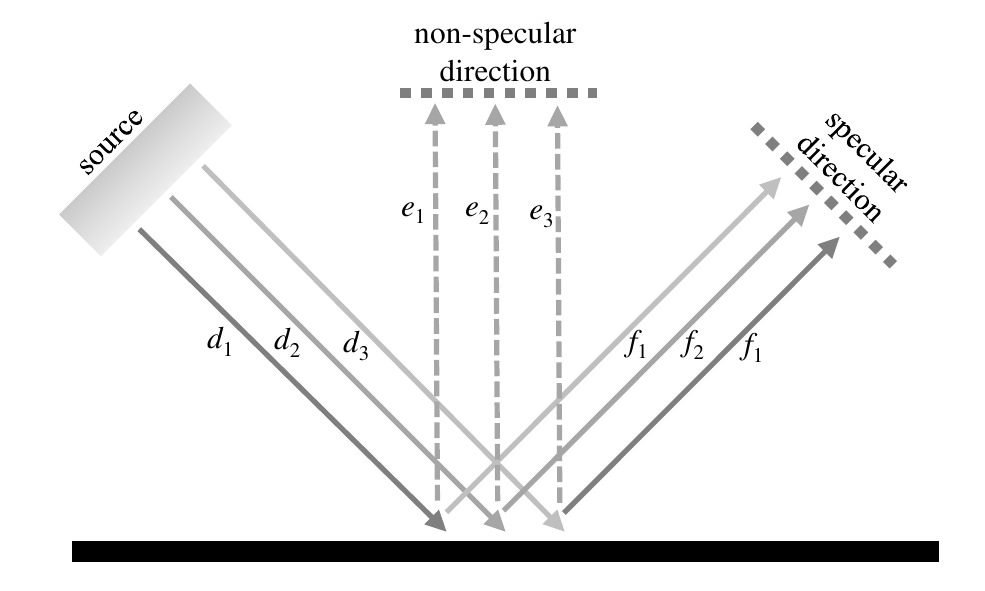}
\caption{Sketch of a wavefront from source to interface, and scattered waves to specular and non-specular directions.}
\label{fig:specular}
\end{figure}

The QSR phenomena from small particles is quite easily explained, starting from the wave-optical mechanism behind the specular reflection. The specular reflection is a simple interference of electromagnetic waves. A sketch of the process is presented in Fig.~\ref{fig:specular}. When the wavefront meets the target the different parts of the wave have traveled different distances, $d_1, d_2$ and $d_3$ in the figure. Generally when the scattered wave is observed at some point, the different parts have traveled distances $d_1+e_1$, $d_2+e_2$, and $d_3+e_3$ which are not equal. Thus, the relative phases of the parts of the wave are different and there is no particular constructive or destructive interference. In specular direction the distances traveled, $d_1+f_1$, $d_2+f_2$ and $d_3+f_3$, are all equal and the relative phases are the same, therefor constructive interference is present. The explanation behind the coherent backscattering (CB) phenomena with the reciprocal ray paths and constructive interference is actually quite close to specular reflection explanation. The differences between QSR and CB are that QSR is single-scattering effect while CB is due to multiple scattering, and that CB is valid for backscattering direction only and QSR for all specular geometries. 

The specular reflection phenomena is usually explained as above with a solid planar interface, but there are no reasons why this mechanism should not work for particles, too. Only requirement is that the particles need to be located in a plane. One could argue that in practice the particles will never be perfectly aligned in a plane, but on the other hand there are no perfectly flat surfaces either. In fact the Bennet-Porteus approximation mentioned earlier shows quite well how the specular power decreases when the surface has small-scale roughness which will effectively introduce more and more phase differences in the specular direction \cite{Penttila-2010-gloss}. While the specular reflection decreases with roughness, the original specular intensity can be magnitudes larger than the diffuse component, so it will remain significant for moderately rough surfaces. The same can be true with particles, so even a small QSR contribution can stand out in the specular direction from the overall scattering pattern.

The existence of a planar or nearly planar layer of particles in the particulate media means that there is some order, some specific structure in the media. In paper coating this is formed by introducing high pressure in the top layer by calender, and in general this can be achieved by compressing the upper layer. Mechanical compression is an obvious mechanism, but also vaporizing or absorbing liquid can introduce capillary forces in the void structure and compress the media.

In what follows I will use electromagnetic wave scattering simulations to study if particulate media can create QSR. Another topic to study here is if the QSR can arise also below the very surface. Wavelength-sized particles can be quite transparent and I have noticed in previous simulations \cite[Sec. 4.1.3 in][]{Penttila-2011-PhD} that ordered structure in the bottom of a finite-depth slab of particles can introduce intensity peak in the specular direction, and I would like to confirm that result.

\subsection*{Light-scattering methods}

To confirm the possible QSR from particulate media I must use exact electromagnetic wave scattering methods that rely on the full treatment of the Maxwell equations. There are phase-dependent interference effects that give raise to the strong signal in the specular direction. Only after the QSR is fully confirmed with exact methods we may think of approximate methods or suitable corrections to approximate methods to deal with the phenomena.

There are a few exact methods and their implementations that could be applied here. These include the discrete dipole approximation (DDA), multiple sphere $T$-matrix (MSTM), and finite difference time domain methods \cite[see e.g.][]{Wriedt-2006-RBC}. From these methods I choose to apply MSTM here, which also dictates that the particulate media discussed here is composed of spheres. In terms of speed and accuracy the MSTM is at least as good as the other methods. The DDA would allow to use any kind of target geometry. It has been already shown that continuous but randomly porous media will produce (quasi-)specular reflection \cite{Penttila-2009-c-coat}, and at this point I do not see that the spherical shapes of the monomers in the media would pose a problem. In favor of the spherical monomers is also the fact that the formation of suitable particulate geometries from spheres is straightforward.

The MSTM and its principles are discussed already in e.g. \cite{Mackowski-2011-MSTM} and I will not repeat that here. I will, however, point out that I will be using the novel feature in this MSTM code, as compared to its predecessor the cluster $T$-matrix code \cite{Mackowski-1996-Tmat}, namely the possibility to use Gaussian beam instead of the plane-wave as the incident field. To study QSR we need plane-parallel media. With exact codes there is always some limit for the number of particles due to the finite computer memory or CPU time in a simulation. It is possible to use MSTM with a computing cluster to simulate a few thousand spheres, results with 3000 spheres, for example, are reported \cite{Mackowski-2011-MSTM}. When building a planar media having that magnitude of wavelength-sized particles, however, the dimensions of the media, i.e. diameter and depth, are quite limited. Therefore it is preferable to minimize the scattering effects arising from the edges of the slab, and this can be done using wave with Gaussian intensity profile that decreases to a low level before the edges.

\subsection*{Is there quasi-specular reflection from particulate media?}

To confirm that there can be QSR from particulate media is the first question to address here. This is done by simulating the intensity phase function from a cylindrical slab of spheres located in the $x,\!y$-plane. The size parameter $x =2 \pi r / \lambda$ for the individual spheres, where $r$ is the radius of the sphere and $\lambda$ is the wavelength, is chosen to be 2. There is no special reason for this particular size parameter, only that I am interested here of roughly wavelength-sized particles since much larger particles would have 'microfacets' that could produce specular scattering in any case. The refractive index for the particles is $1.31 + i 0$ throughout the article.

\begin{figure}
\centering
\includegraphics{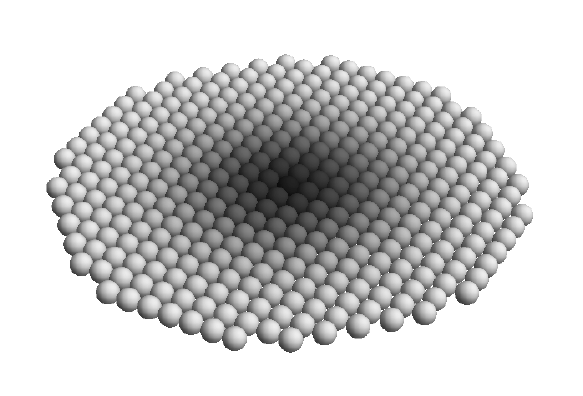}
\caption{Layer of spheres in circular area of regular lattice. The gray-scale value of individual sphere is comparable to its independent extinction efficiency, induced by the Gaussian incident beam. Higher efficiencies show black and gray shades, and lower efficiencies whiter shades. I will refer to this layer as \emph{type 1} in the text.}
\label{fig:gaussian}
\end{figure}

For the size of the cylindrical layer of spheres I will consider a regular lattice in $(x,y)$ that has sidelength of 84 size parameter units (spu, physical length $\! \times \, 2 \pi / \lambda$). This lattice could hold 441 spheres with $x=2$, but I will cut a circular area from this lattice which will reduce the number of spheres in the layer to 349. This layer can be seen in Fig.~\ref{fig:gaussian}. The Gaussian beam intensity profile is selected so that the so-called beam constant is $0.03$, meaning that the intensity of the beam will drop as $\exp(-(0.03 \, d)^2)$, where $d$ is the distance (in spu) from the lattice center.

The direction of the possible QSR is, naturally, dependent on the direction of the incident light. I will use the zenith angle, azimuth angle and the principal plane to define the directions. Zenith angle is measured from the layer's normal vector (i.e., the $z$-axis) and azimuth angle is the projected angle around the normal in the $x,\!y$-plane. The so-called principal plane is the plane defined by the incident and zenith directions. Let us agree that the specular direction, which is always in the principal plane, has a positive zenith angle $\theta_s$, and the incident direction $\theta_i$ a negative value, $-\theta_i = \theta_s$. In zenith illumination $\theta_i = 0$ and the azimuth angles of the incident direction, specular direction and principal plane are not uniquely defined.

From the practical point of view this zenith illumination geometry is preferred in simulations. With many scattering codes, including fixed orientation MSTM and many DDA codes, the scattering can be computed efficiently over different scattering planes. Scattering planes are planes around the incident direction vector. When $\theta_i = 0$ all the scattering planes are principal planes containing the specular direction, but with different azimuth angles. On the other hand, we need to average the scattering over different azimuthal orientations of the layer. With zenith illumination this averaging can be done by averaging efficiently over scattering planes, otherwise averaging must be done using principal planes with different azimuth angles each requiring independent simulation, thus using much more computing time. In this study the orientation over azimuth was done with one degree steps.

There are possible caveats in using this zenith illumination geometry. Other special scattering phenomena can arise, namely the coherent backscattering and the shadow-hiding, which both can increase the intensity in the backscattering geometry \cite[e.g.][]{Hapke-1986-opposition,Shkuratov-1999-Clementine,Shkuratov-2002-analogs,Mishchenko-2006-rtetcoba,Muinonen-2012-CBRT}. Therefore we must first make sure that the QSR exists with other than zenith illumination geometry.

\begin{figure}
\centering
\includegraphics{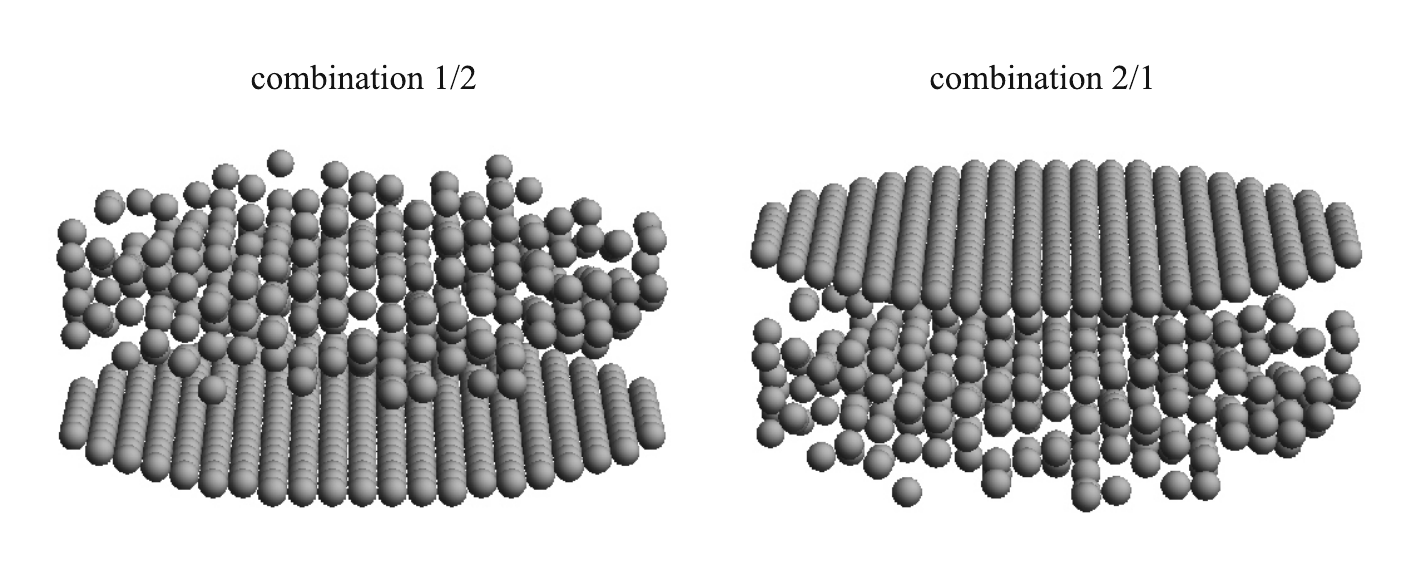}
\caption{Example of combinations 1/2 and 2/1 with type 1 (planar) layer and type 2 layer (random $z$-coordinate).}
\label{fig:lay12}
\end{figure}

Other issue that we must take into account is that we are assuming that QSR will happen for planar layer of particles because of the interference behind the specular phenomena. In the same manner we must assume that the QSR will not arise for a particulate layer that is not planar. This is tested with three realizations of similar setup of circular area of regular lattice of spheres, but where the $z$-coordinate of the individual spheres is uniformly distributed random variable between -2 and 2 (spu). See Fig.~\ref{fig:lay12} for an example of both planar layer (type 1) and layer with random $z$-coordinate (type 2).

\begin{figure}
\centering
\includegraphics{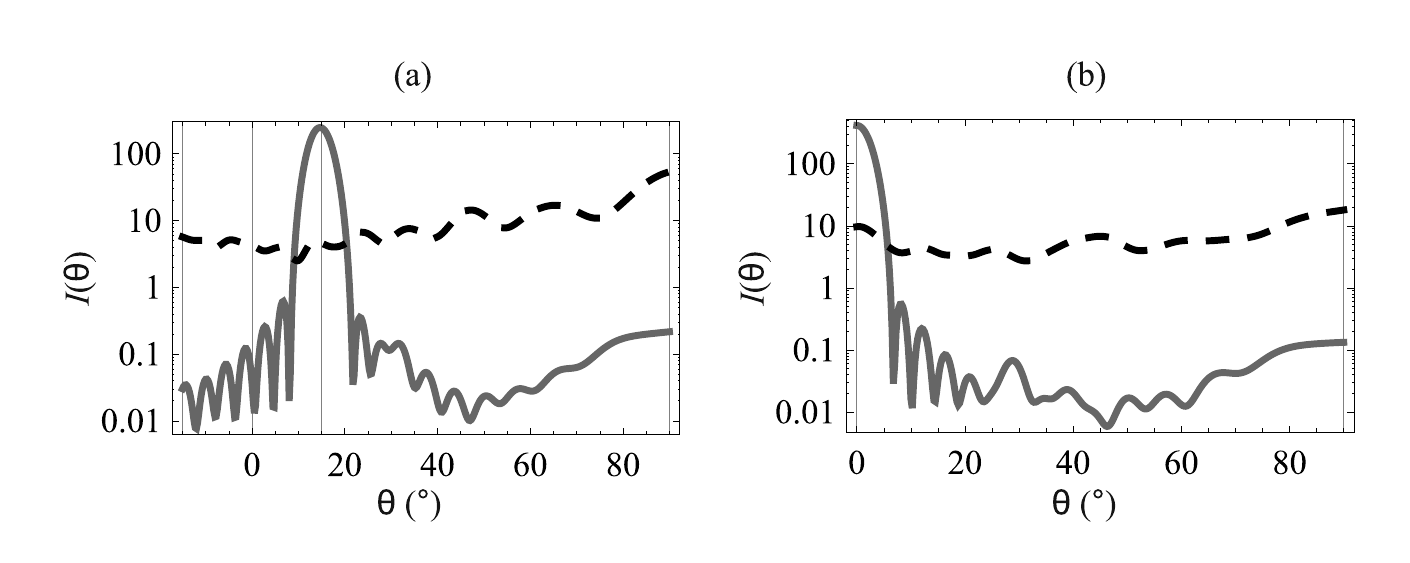}
\caption{Intensity of the scattered light as a function of the zenith angle $\theta$, i.e., angle measured form the surface normal. Incident light has angle $\theta=-15\degr$ with the layer normal in case (a) on the left, and angle $\theta=0\degr$ in case (b) on the right. The gray solid line is for the type 1 planar layer, and the black dashed line for type 2 layers, averaged over three realizations. Only the reflected scattered light is shown here, not the transmitted light through the layer.}
\label{fig:zenith}
\end{figure}

The results, i.e. the scattered directional intensity for type 1 layer and type 2 layers, averaged over three realizations, are shown in Fig.~\ref{fig:zenith}(a). We can clearly see that when incident light has $\theta=-15\degr$ there is a distinct peak at $\theta=15\degr$ with type 1 planar layer. The type 2 layers show very flat intensity profile around $\theta=15\degr$. It is indisputably shown that there is a strong QSR effect with the planar type 1 layer. The intensity at specular direction is more than 50 times stronger with the planar layer.

Now that it has been shown that QSR can exist I will show in Fig.~\ref{fig:zenith}(b) that it can be studied here in the more practical backscattering illumination geometry. The overall profile is the same -- type 1 layer produces clear specular peak that is about 40 times stronger than the flat profile of the type 2 layers. It is a fair assumption that we can use backscattering setup here to study the QSR. We can estimate the possible contribution from other backscattering effects by comparing the intensity values between specular direction ($\theta_s = 15\degr$) in Fig.~\ref{fig:zenith}(a) and specular/backscattering direction ($\theta_s = 0\degr$) in Fig.~\ref{fig:zenith}(b). For type 1 layer the intensity is enhanced by a factor of $1.7$ and for type layer by a factor of $2.1$ when moving from $\theta_s = 15\degr$ to $\theta_s = 0\degr$. The theoretical maximum contribution from coherent backscattering is two, and these simulated values are quite nicely in that neighbourhood \cite{Mishchenko-2006-rtetcoba}. If we compare the strengths of the possible backscattering effects ($\approx$ 2) and the QSR effects ($\approx$ 40--50) we can see that it is quite safe to discard the other backscattering effects and use zenith illumination when studying QSR. I will return this subject of QSR and other backscattering effects in Sect.~Discussion.

\subsection*{Is the effect coming from the topmost layer only?}

\begin{figure}
\centering
\includegraphics{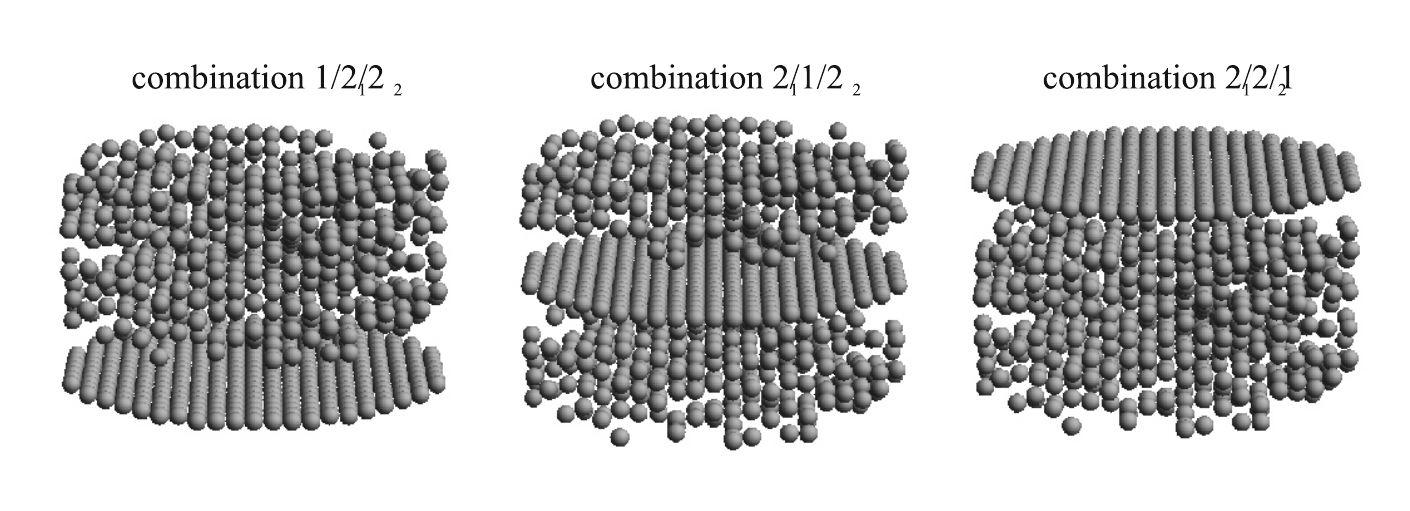}
\caption{Example of combinations $1/2_1/2_2$, $2_1/1/2_2$, $2_1/2_2/1$ with type 1 layer and type 2 layers. I use three different realizations of type 2 layers, $2_1, 2_2$, and $2_3$.}
\label{fig:lay123}
\end{figure}

It is quite possible to assume that if and when this QSR can exist, it will arise only from the very topmost particles and if there are particles above the planar structure it will destruct the possible QSR. I will test this by stacking up these type 1 and 2 layers and computing their scattering properties. This test is executed using two- or three-layer combinations of the elementary layers. The layers are combined so that there is a small gap between the layers and they do not intersect, see examples in Figs.~\ref{fig:lay12} and \ref{fig:lay123}.

\begin{figure}
\centering
\includegraphics{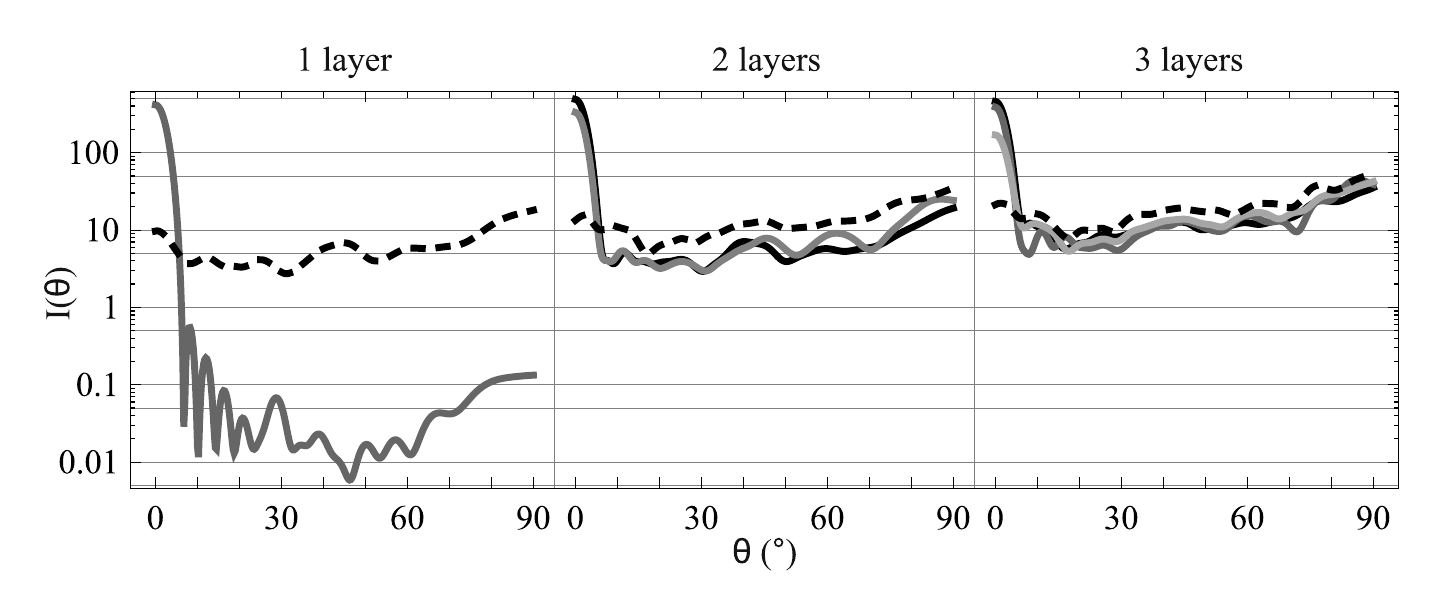}
\caption{Reflected intensity for different combinations of type 1 and 2 layers as a function of the zenith angle $\theta$. In all the subfigures the black dashed line is used for the group without type 1 layer. In the two- and three-layer subfigures the solid line transforms lighter as the one type 1 layer in the slab moves down from the surface.}
\label{fig:Igroup}
\end{figure}

There are 12 possible two-layer combinations of one type 1 and three type 2 layers, and 24 three-layer combinations. The two-layer combinations can be logically divided into three groups: combinations where type 1 layer is at bottom, combinations where type 1 layer is at top, and combinations where there is no type 1 layer. With three layers we can find four groups: type 1 at bottom, in center, at top, or no type 1 layer. The scattering results from these groups, seen in Fig.~\ref{fig:Igroup}, show quite interesting pattern. All the groups where there is type 1 planar layer present will produce QSR. The strength of QSR is smallest with combinations where type 1 layer is at bottom, but still very evident. This is a strong proof of the fact that at least a moderate number of obstructing particles will not destroy the QSR.

\subsection*{Is the effect linked to specific uni-sized particles?} 

\begin{figure}
\centering
\includegraphics{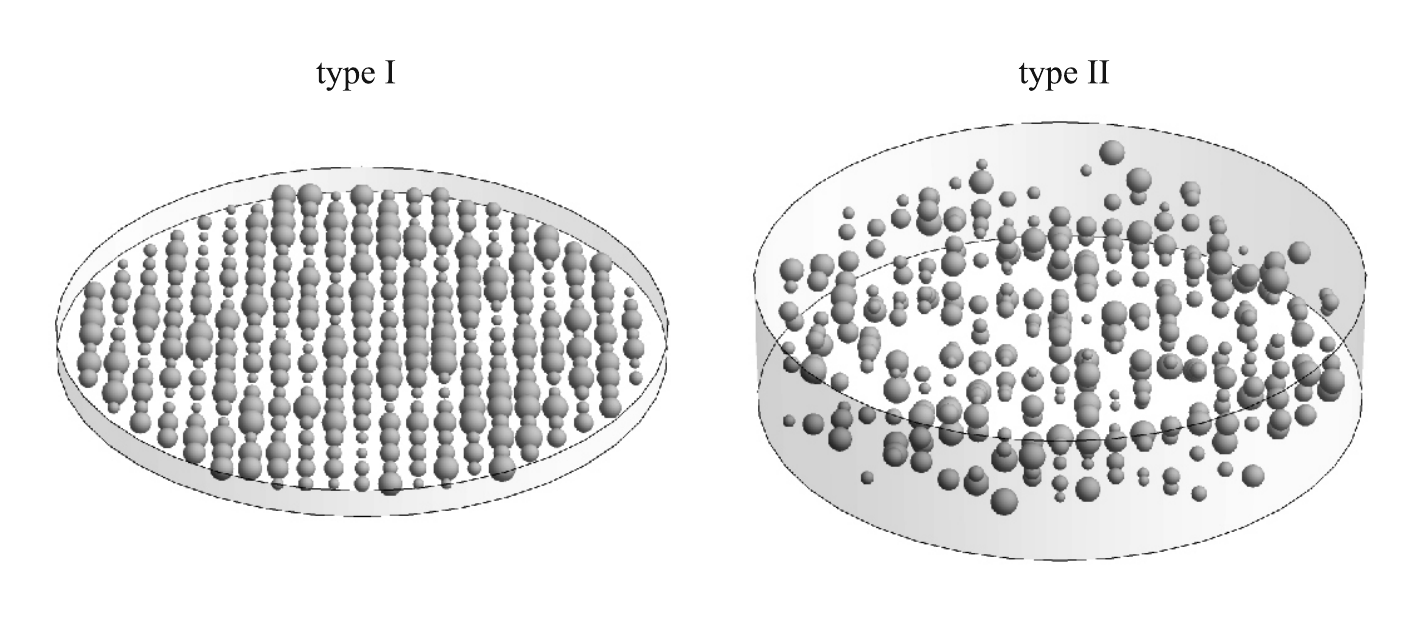}
\caption{Type I layer on the left, and one of three realizations of type II layer on the right. The size parameters for the particles are drawn from uniform distribution between 1 and 2.75.}
\label{fig:layIetII}
\end{figure}

One might speculate that the QSR is happening above only because I am simulating scattering from a collection of particles that all have perfectly the same size and shape, which will enable the coherent interference. Again, this can be tested. I will simulate the same results as above --- the scattering for different number of layers and for different combinations --- but I will use type I and type II layers in the combinations. The basic setup for I and II is the same as for type 1 and 2 layers, so there is a circular area of the regular lattice for the $(x,y)$-coordinates of the particles, but this time the particles have a size distribution. The size parameter $x$ can vary between 1 and $x_{max}$ with uniform distribution whereas previously $x$ was 2 for all the spheres. The upper limit for the size parameter, $x_{max}$, is set here so that the expected volume of the random-sized sphere is the same as for sphere with $x=2$. It turns out that this value for $x_{max}$ is about $2.75$. Naturally this means that the lattice must be enlarged to accommodate spheres with maximum size of $2.75$, thus the lattice will have a sidelength of about 115 spu. Example of type I and II layers is shown in Fig.~\ref{fig:layIetII}.

\begin{figure}
\centering
\includegraphics{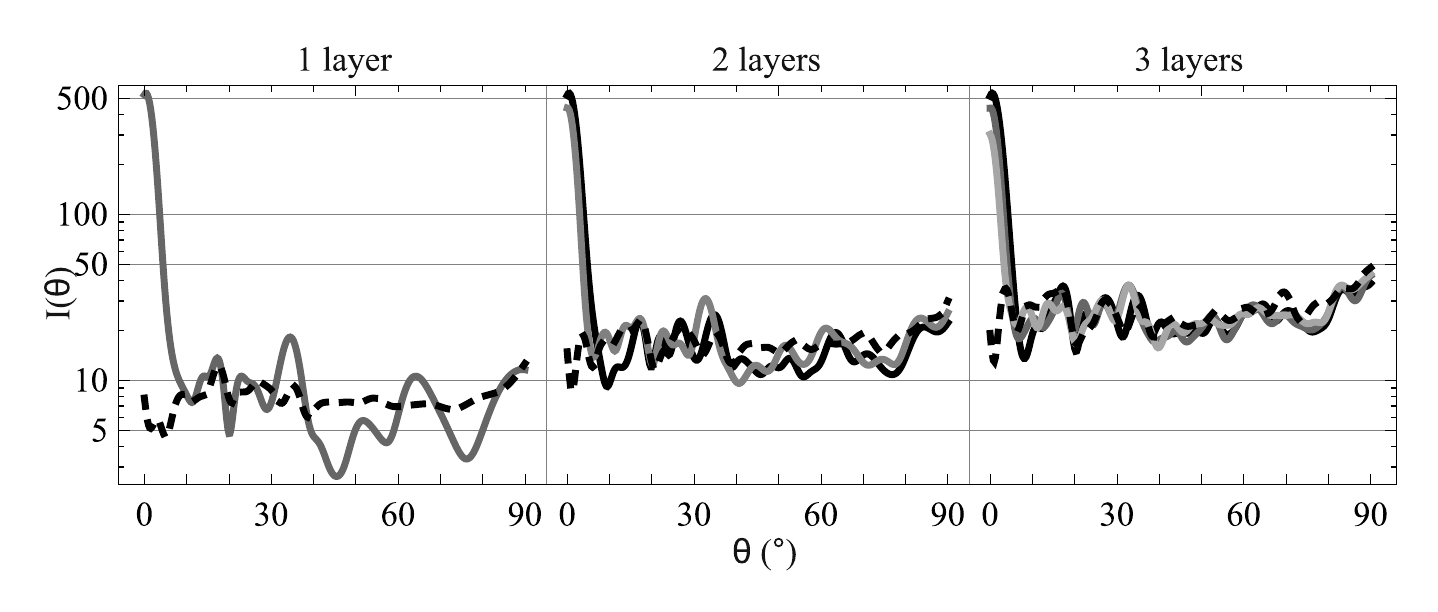}
\caption{Reflected intensity for different combinations of type I and II layers as a function of the zenith angle $\theta$. In all the subfigures the black dashed line is used for the group without type I layer. In the two- and three-layer subfigures the solid line transforms lighter as the one type I layer in the slab moves down from the surface.}
\label{fig:Irgroup}
\end{figure}

Using the same two- and three-layer combinations of type I and II layers we can again see in Fig.~\ref{fig:Irgroup} a very strong difference between combinations without planar (type I) layer and combinations where type I is present somewhere in the slab. Actually it seems that this difference is even stronger than with uni-sized particles. It is therefore concluded that same-sized constituents in particulate material are not required for the QSR to arise.

\subsection*{Is the effect due to the regular lattice of particles?}

\begin{figure}
\centering
\includegraphics{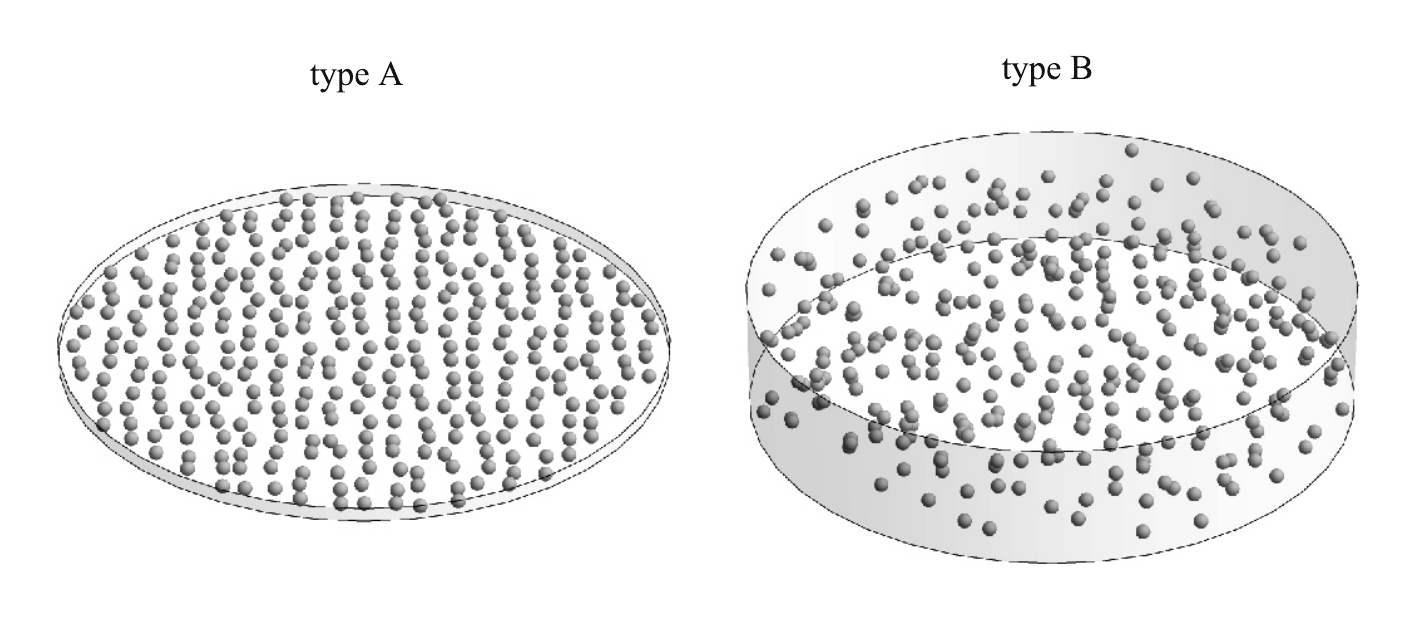}
\caption{Type A layer on the left, and one of three realizations of type B layer on the right. The lattice cell is larger than the particle and the position of the particle in that cell is random.}
\label{fig:layAetB}
\end{figure}

One more suspicion behind the practical meaningfulness of the QSR in the abovementioned cases is that maybe the effect only arises from the regular lattice where the particles are located. This must be checked, and I will use again new layer types (A and B) to tackle this issue. In these types I will use regular lattice where the area of a single cell is larger than the area needed for the particle, so that the position of the particle in the single lattice cell can be random. I will set the sidelength of the cell to 8 spu and use spheres with $x=2$, so the diameter of the sphere is 4 spu. Each sphere is positioned randomly somewhere inside the corresponding cell. The sidelength for the whole grid will be 168 spu. Examples of the type A and B layers are shown in Fig.~\ref{fig:layAetB}.

\begin{figure}
\centering
\includegraphics{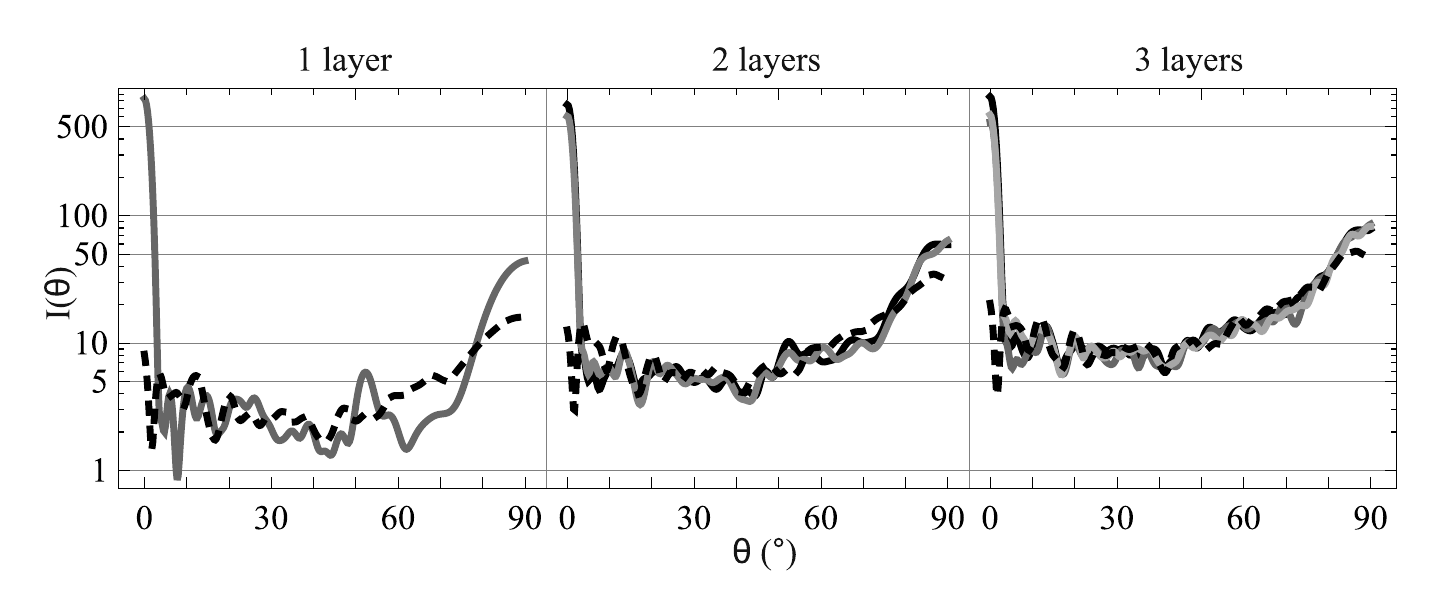}
\caption{Reflected intensity for different combinations of type A and B layers as a function of zenith angle $\theta$. In all the subfigures the black dashed line is used for the group without type A layer. In the two- and three-layer subfigures the solid line transforms lighter as the one type A layer in the slab moves down from the surface.}
\label{fig:Irxygroup}
\end{figure}

Once again it can be seen from Fig.~\ref{fig:Irxygroup} that the random locations of the particles in the planar layer will not destroy the QSR. On the contrary, the difference between slabs with and without the planar layer is even more prominent.

\section*{Discussion}
\label{sec:disc}

I have shown here that there can exist a strong specular-type intensity peak from particulate media if it includes a planar layer of particles. This QSR will arise even if the planar layer is not on the very top of the media, does not consist of same-sized particles, or is not located in regular lattice. It seems that the only requirement is the same as for specular reflection from solid surfaces -- the (particulate) media has to have such a structure that there is planar arrangement with the particles.

\begin{figure}
\centering
\includegraphics{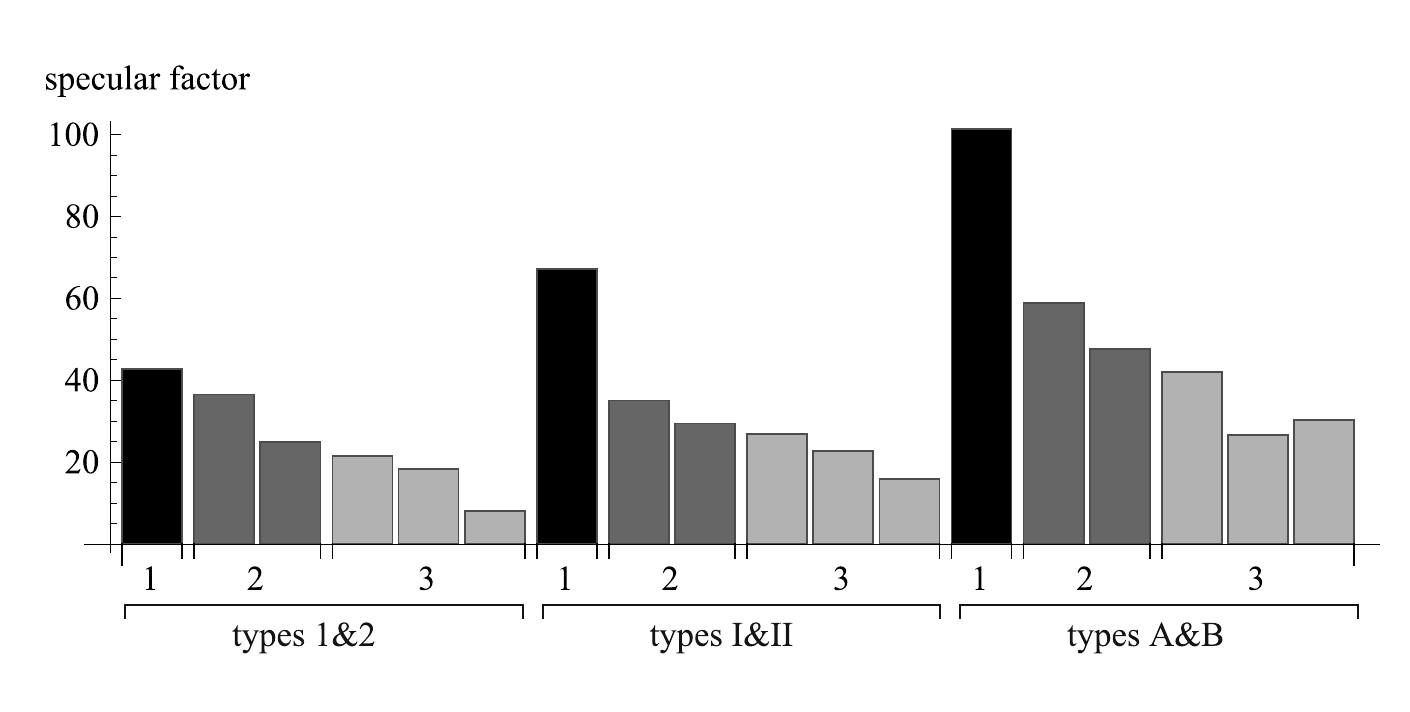}
\caption{Combined specular factors for all the types (1\&2, I\&II, A\&B) and their one-, two-, or three-layer combinations. The specular factor is the enhancement in the intensity at specular direction, relative to the non-specular case. The two- and three-layer combinations in the plot are arranged so that the first bar corresponds to the case where the planar layer is at the top of the slab, and the last bar to the case where it is at the bottom.}
\label{fig:gfactor}
\end{figure}

The strength of this QSR can be quite high. In Fig.~\ref{fig:gfactor} I present a summary of the scattering results in this study regarding the specular direction with the so-called specular factor. This is the intensity in the specular direction, relative to that of similar slab without the planar layer. Quite interestingly the factor is actually lowest with the case of regular lattice and same-sized particles (type 1 and 2 layers). In any case, the specular factor can have strengths from 8 to 40 even when there are two layers of obstructing particles on top of the planar layer.

The practical consequences of the confirmed QSR from particulate media awaits for more studies on the subject. When modeling the observations from natural targets we need to take into account the variability that is always present. There will not exist perfect planar layers of particles, and the particles will have distribution in both size and shape. The possible magnitude of QSR in observations would need further simulations where the perfect planar layer is disturbed in the $z$-direction and where random non-spherical shapes are considered. In the more theoretical side we would benefit from near-field simulations or finite difference time domain --type simulations where we could see in detailed level the QSR arising from the particles.

Despite the need for further studies, one might already take the possibility of QSR in certain situations into account. Especially if we are studying backscattering effects either by simulations or by laboratory measurements from the surface of plane-parallel particulate media in zenith illumination geometry, we might end up in a situation where all the three possible effects (coherent backscattering, shadow-hiding and QSR) are contributing. In such a case it is hard to separate afterward the roles the different mechanisms play in the backscattering enhancement. Fortunately it is easy to avoid such entanglement of all three effects --- one needs to use non-zenith illumination geometry. If the illumination is from some angle $\theta$, the coherent backscattering should be seen at $\theta$ and the QSR at $-\theta$. Thus, when $\theta$ is non-zero, coherent backscattering and QSR would be seen in separate directions. Besides, at $\theta \neq 0$, the shadow-hiding contribution is negligible.

\section*{Acknowledgments}

Kari Lumme is acknowledged for his input on this work and for fruitful discussions. Michael Mishchenko is thanked for the comments on this subject when acting as an opponent in my PhD defense. The computations presented have been made using CSC computing resources. CSC is the Finnish IT center for science and is owned by the Ministry of Education. This project has been funded by the Finnish Academy grant no. 127461.


\begin{thebibliography}{16}
\providecommand{\natexlab}[1]{#1}
\providecommand{\url}[1]{\texttt{#1}}
\providecommand{\urlprefix}{URL }
\expandafter\ifx\csname urlstyle\endcsname\relax
  \providecommand{\doi}[1]{doi:\discretionary{}{}{}#1}\else
  \providecommand{\doi}[1]{doi:\discretionary{}{}{}\begingroup
  \urlstyle{rm}\url{#1}\endgroup}\fi
\providecommand{\bibinfo}[2]{#2}

\bibitem[{Torrance and Sparrow(1967)}]{Torrance-1967-specular}
\bibinfo{author}{K.~Torrance}, \bibinfo{author}{E.~Sparrow},
  \bibinfo{title}{Theory for off-specular reflection from roughened surfaces},
  \bibinfo{journal}{Journal of Optical Society of America} \bibinfo{volume}{57}
  (\bibinfo{year}{1967}) \bibinfo{pages}{1105--1114}.

\bibitem[{Beckmann and Spizzichino(1963)}]{Beckmann-1963-surfaces}
\bibinfo{author}{P.~Beckmann}, \bibinfo{author}{A.~Spizzichino},
  \bibinfo{title}{The scattering of electromagnetic waves from rough surfaces},
  \bibinfo{publisher}{Pergamon}, \bibinfo{address}{New York},
  \bibinfo{year}{1963}.

\bibitem[{Borovoi et~al.(2012)Borovoi, Konoshonkin, and
  Kolokolova}]{Borovoi-2012-Glints}
\bibinfo{author}{A.~Borovoi}, \bibinfo{author}{A.~Konoshonkin},
  \bibinfo{author}{L.~Kolokolova}, \bibinfo{title}{Glints from particulate
  media and wavy surfaces}, \bibinfo{journal}{Journal of Quantitative
  Spectroscopy \& Radiative Transfer}
  \bibinfo{volume}{113}~(\bibinfo{number}{18}) (\bibinfo{year}{2012})
  \bibinfo{pages}{2542--2551}.

\bibitem[{Janssen et~al.(2011)Janssen, Le~Gall, and Wye}]{Janssen-2011-Titan}
\bibinfo{author}{M.~Janssen}, \bibinfo{author}{A.~Le~Gall},
  \bibinfo{author}{L.~Wye}, \bibinfo{title}{Anomalous radar backscatter from
  {T}itan's surface?}, \bibinfo{journal}{Icarus}
  \bibinfo{volume}{212}~(\bibinfo{number}{1}) (\bibinfo{year}{2011})
  \bibinfo{pages}{321--328}.

\bibitem[{Soderblom et~al.(2012)Soderblom, Barnes, Soderblom, Brown, Griffith,
  Nicholson, Stephan, Jaumann, Sotin, Baines, Buratti, and
  Clark}]{Soderblom-2012-specular}
\bibinfo{author}{J.~Soderblom}, \bibinfo{author}{J.~Barnes},
  \bibinfo{author}{L.~Soderblom}, \bibinfo{author}{R.~Brown},
  \bibinfo{author}{C.~Griffith}, \bibinfo{author}{P.~Nicholson},
  \bibinfo{author}{K.~Stephan}, \bibinfo{author}{R.~Jaumann},
  \bibinfo{author}{C.~Sotin}, \bibinfo{author}{K.~Baines},
  \bibinfo{author}{B.~Buratti}, \bibinfo{author}{R.~Clark},
  \bibinfo{title}{Modeling specular reflections from hydrocarbon lakes on
  {T}itan}, \bibinfo{journal}{Icarus}
  \bibinfo{volume}{220}~(\bibinfo{number}{2}) (\bibinfo{year}{2012})
  \bibinfo{pages}{744--751}.

\bibitem[{Penttil\"a and Lumme(2010)}]{Penttila-2010-gloss}
\bibinfo{author}{A.~Penttil\"a}, \bibinfo{author}{K.~Lumme},
  \bibinfo{title}{Specular gloss simulations of media with small-scale
  roughness}, in: \bibinfo{editor}{K.~Muinonen},
  \bibinfo{editor}{A.~Penttil\"a}, \bibinfo{editor}{H.~Lindqvist},
  \bibinfo{editor}{T.~Nousiainen}, \bibinfo{editor}{G.~Videen} (Eds.),
  \bibinfo{booktitle}{Proceedings of the 12th Conference on Electromagnetic and
  Light Scattering}, \bibinfo{address}{Helsinki, Finland},
  \bibinfo{pages}{230--233}, \bibinfo{year}{2010}.

\bibitem[{Penttil\"a(2011)}]{Penttila-2011-PhD}
\bibinfo{author}{A.~Penttil\"a}, \bibinfo{title}{Light scattering in random
  media with wavelength-scale structures: astronomical and industrial
  applications}, \bibinfo{type}{Doctoral thesis}, \bibinfo{school}{University
  of Helsinki, Finland},
  \urlprefix\url{http://urn.fi/URN:ISBN:978-952-10-7073-0},
  \bibinfo{year}{2011}.

\bibitem[{Wriedt et~al.(2006)Wriedt, Hellmers, Eremina, and
  Schuh}]{Wriedt-2006-RBC}
\bibinfo{author}{T.~Wriedt}, \bibinfo{author}{J.~Hellmers},
  \bibinfo{author}{E.~Eremina}, \bibinfo{author}{R.~Schuh},
  \bibinfo{title}{Light scattering by single erythrocyte: {C}omparison of
  different methods}, \bibinfo{journal}{Journal of Quantitative Spectroscopy \&
  Radiative Transfer} \bibinfo{volume}{100}~(\bibinfo{number}{1--3})
  (\bibinfo{year}{2006}) \bibinfo{pages}{444--456}.

\bibitem[{Penttil\"a and Lumme(2009)}]{Penttila-2009-c-coat}
\bibinfo{author}{A.~Penttil\"a}, \bibinfo{author}{K.~Lumme},
  \bibinfo{title}{The effect of the properties of porous media on light
  scattering}, \bibinfo{journal}{Journal of Quantitative Spectroscopy \&
  Radiative Transfer} \bibinfo{volume}{110}~(\bibinfo{number}{18})
  (\bibinfo{year}{2009}) \bibinfo{pages}{1993--2001}.

\bibitem[{Mackowski and Mishchenko(2011)}]{Mackowski-2011-MSTM}
\bibinfo{author}{D.~Mackowski}, \bibinfo{author}{M.~Mishchenko},
  \bibinfo{title}{A multiple sphere {$T$}-matrix {F}ortran code for use on
  parallel computer clusters}, \bibinfo{journal}{Journal of Quantitative
  Spectroscopy \& Radiative Transfer}
  \bibinfo{volume}{112}~(\bibinfo{number}{13}) (\bibinfo{year}{2011})
  \bibinfo{pages}{2182--2192}.

\bibitem[{Mackowski and Mishchenko(1996)}]{Mackowski-1996-Tmat}
\bibinfo{author}{D.~Mackowski}, \bibinfo{author}{M.~Mishchenko},
  \bibinfo{title}{Calculation of the {$T$}-matrix and the scattering matrix for
  ensembles of spheres}, \bibinfo{journal}{Journal of Optical Society of
  America A} \bibinfo{volume}{13} (\bibinfo{year}{1996})
  \bibinfo{pages}{2266--2278}.

\bibitem[{Hapke(1986)}]{Hapke-1986-opposition}
\bibinfo{author}{B.~Hapke}, \bibinfo{title}{Bidirectional reflectance
  spectroscopy : 4. {T}he extinction coefficient and the opposition effect},
  \bibinfo{journal}{Icarus} \bibinfo{volume}{67}~(\bibinfo{number}{2})
  (\bibinfo{year}{1986}) \bibinfo{pages}{264--280}.

\bibitem[{Shkuratov et~al.(1999)Shkuratov, Kreslavsky, Ovcharenko, Stankevich,
  Zubko, Pieters, and Arnold}]{Shkuratov-1999-Clementine}
\bibinfo{author}{Y.~Shkuratov}, \bibinfo{author}{M.~Kreslavsky},
  \bibinfo{author}{A.~Ovcharenko}, \bibinfo{author}{D.~Stankevich},
  \bibinfo{author}{E.~Zubko}, \bibinfo{author}{C.~Pieters},
  \bibinfo{author}{G.~Arnold}, \bibinfo{title}{Opposition Effect from
  {C}lementine Data and Mechanisms of Backscatter}, \bibinfo{journal}{Icarus}
  \bibinfo{volume}{141}~(\bibinfo{number}{1}) (\bibinfo{year}{1999})
  \bibinfo{pages}{132--155}.

\bibitem[{Shkuratov et~al.(2002)Shkuratov, Ovcharenko, Zubko, Miloslavskaya,
  Muinonen, Piironen, Nelson, Smythe, Rosenbush, and
  Helfenstein}]{Shkuratov-2002-analogs}
\bibinfo{author}{Y.~Shkuratov}, \bibinfo{author}{A.~Ovcharenko},
  \bibinfo{author}{E.~Zubko}, \bibinfo{author}{O.~Miloslavskaya},
  \bibinfo{author}{K.~Muinonen}, \bibinfo{author}{J.~Piironen},
  \bibinfo{author}{R.~Nelson}, \bibinfo{author}{W.~Smythe},
  \bibinfo{author}{V.~Rosenbush}, \bibinfo{author}{P.~Helfenstein},
  \bibinfo{title}{The Opposition Effect and Negative Polarization of Structural
  Analogs for Planetary Regoliths}, \bibinfo{journal}{Icarus}
  \bibinfo{volume}{159} (\bibinfo{year}{2002}) \bibinfo{pages}{396--416}.

\bibitem[{Mishchenko et~al.(2006)Mishchenko, Travis, and
  Lacis}]{Mishchenko-2006-rtetcoba}
\bibinfo{author}{M.~Mishchenko}, \bibinfo{author}{L.~Travis},
  \bibinfo{author}{A.~Lacis}, \bibinfo{title}{Multiple scattering of light by
  particles. {R}adiative transfer and coherent backscattering},
  \bibinfo{publisher}{Cambridge University Press}, \bibinfo{year}{2006}.

\bibitem[{Muinonen et~al.(2012)Muinonen, Mishchenko, Dlugach, Zubko,
  Penttil\"a, and Videen}]{Muinonen-2012-CBRT}
\bibinfo{author}{K.~Muinonen}, \bibinfo{author}{M.~Mishchenko},
  \bibinfo{author}{J.~Dlugach}, \bibinfo{author}{E.~Zubko},
  \bibinfo{author}{A.~Penttil\"a}, \bibinfo{author}{G.~Videen},
  \bibinfo{title}{Coherent backscattering verified numerically for a finite
  volume of spherical particles}, \bibinfo{journal}{Astrophysical Journal}
  \bibinfo{volume}{760}~(\bibinfo{number}{2}) (\bibinfo{year}{2012})
  \bibinfo{pages}{118--128}.

\end{thebibliography}
\end{document}